\documentclass[preprint,eqsecnum,aps]{revtex4}
\usepackage{psfig}
\usepackage{graphicx}%
\usepackage{dcolumn}
\usepackage{amsmath}

\makeatletter
\def\btt#1{\texttt{\@backslashchar#1}}%
\DeclareRobustCommand\bblash{\btt{\@backslashchar}}%
\makeatother

\begin{document}

\title{Attractor Phantom Solution}

\author{Jian-gang Hao}
\author{Xin-zhou Li}\email{kychz@shtu.edu.cn}
\affiliation{SUCA, Shanghai United Center for Astrophysics,
Shanghai Normal University, 100 Guilin Road, Shanghai 200234,China
}%

\date{\today}

\begin{abstract}
ABSTRACT: In light of recent study on the dark energy models that
manifest an equation of state $w<-1$, we investigate the
cosmological evolution of such a phantom field in a specific
potential, exponential potential in this paper. The phase plane
analysis show that the there is a late time attractor solution in
this model, which address the similar issues as that of fine
tuning problems in conventional quintessence models. The equation
of state $w$ is determined by the attractor which is dependent on
the $\lambda$ parameter in the potential. We also show that this
model is stable for our present observable Universe.
\end{abstract}

\maketitle

\vspace{0.4cm} \noindent \textbf{1. Introduction} \vspace{0.4cm}

Recent observation shows that our universe is made up of roughly
two third of dark energy that has negative pressure and can drive
the accelerating expansion of the universe\cite{newobservation}.
Present data from the observation allows the equation of state in
the range $-1.62<w<-0.74$\cite{melchiorri}. However, the equation
of state of conventional quintessence models\cite{steinhardt}that
based on a scalar field and positive kinetic energy can not evolve
to the the regime of $w<-1$. Some
authors\cite{caldwell1,sahni,parker,chiba,boisseau,schulz,faraoni,maor,onemli,
torres,carroll,frampton,caldwell2,gibbons} investigated a phantom
field model which has negative kinetic energy and can realize the
$w<-1$ in its evolution. Although the introduction of a phantom
field causes many theoretical problems such as the violation of
some widely accepted energy condition and the rapid vacuum
decay\cite{carroll}, it is still very interesting in the sense
that it can fit current observations well. Comparing with other
approaches to realizing the $w<-1$, such as the modification of
Friedman equation, it seems more economical.

In this paper, provisionally leaving aside the theoretical puzzles
about phantom field, we study the detailed evolution of the
phantom field and the attractor property of its solution via a
specific model. by using the qualitative approach to the dynamical
system, phase plane analysis, we prove the existence of a late
time attractor solution, at which the phantom become dominant and
the equation of state is freezed with only small oscillation.
Exponential potentials attract much attention because they can be
derived from the effective interaction in string theory, higher
dimensional gravity and the non-perturbative effects such as
gaugino condensation\cite{string} and their roles in cosmological
context have also been widely investigated\cite{lucchin}. We
therefore consider the phantom evolution in exponential potential
in the following.

\vspace{0.4cm} \noindent\textbf{2. Phantom in Exponential
Potential}
 \vspace{0.4cm}

In the spatially flat Robertson-Walker metric,
\begin{equation}
ds^{2}=dt^{2}-a^{2}(t)(dx^{2}+dy^{2}+dz^{2})
\end{equation}

\noindent The Lagrangian density for the spatially homogeneous
phantom field is

\begin{equation}
L_{\Phi}=-\frac{1}{2}g^{\mu\nu}(\partial_{\mu}\phi)(\partial_{\nu}\phi)-V(\phi)
\end{equation}

\noindent when consider the presence of baryotropic fluid, the
action for the model is

\begin{equation}
S=\int
d^{4}x\sqrt{-g}(-\frac{1}{2\kappa^2}R_{s}-p_{\gamma}+L_{\phi})
\end{equation}

\noindent where $g$ is the determinant of the metric tensor $
g_{\mu\nu}$, $\kappa^2=8\pi G$, $R_{s}$ is the Ricci scalar,and
$\rho_{\gamma}$ is the density of fluid with a baryotropic
equation of state $p_{\gamma}=(\gamma-1)\rho_{\gamma}$, where
$0\leq \gamma\leq2$ is a constant that relates to the equation of
state by $w=\gamma-1$. By varying the action, one can obtain the
Einstein equations and the equations of motion for the scalar
field as

\begin{equation}\label{sys1}
\dot{H}=-\frac{\kappa^2}{2}(\rho_\gamma+p_\gamma-\dot{\phi}^2)
\end{equation}

\begin{equation}\label{sys2}
\dot{\rho_\gamma}=-3H(\rho_\gamma+p_\gamma)
\end{equation}

\begin{equation}\label{sys3}
\ddot{\phi}+3H\dot{\phi}+\lambda\kappa V(\phi)=0
\end{equation}

\begin{equation}\label{sys4}
H^2=\frac{\kappa^2}{3}[\rho_\gamma+\rho_{\Phi}]
\end{equation}

\noindent where

\begin{equation}\label{rho11}
\rho_{\Phi}=-\frac{1}{2}\dot{\phi}^{2}+V(\phi)
\end{equation}

\begin{equation}\label{p11}
p_{\Phi}=-\frac{1}{2}\dot{\phi}^{2}-V(\phi)
\end{equation}

\noindent are the energy density and pressure of the $\Phi$ field
respectively , and $H$ is Hubble parameter. The potential
$V(\phi)$ is exponentially dependent on $\phi$ as
$V(\phi)=V_0\exp(-\lambda\kappa \phi)$.The equation-of-state
parameter for the phantom is

\begin{equation}
w =\frac{p_\Phi}{\rho_\Phi}=
\frac{\dot{\phi}^{2}+2V(\phi)}{\dot{\phi}^{2}-2V(\phi)}
\end{equation}

It is clear that the phantom field could realize the equation of
state $w<-1$ when

\begin{equation}
0<\dot{\phi}^{2}<2V(\phi)
\end{equation}

\vspace{0.4cm} \noindent \textbf{3. Attractor Solution of Phantom
Model} \vspace{0.4cm}

In this section, we investigate the global structure of the
dynamical system via phase plane analysis and compute the
cosmological evolution by numerical analysis. Introduce the
following variables

\begin{eqnarray}\label{vari}
x=&&\frac{\kappa}{\sqrt{6}H}\dot{R}\\\nonumber
y=&&\frac{\kappa\sqrt{V(R)}}{\sqrt{3}H}\\\nonumber
N=&&\log a
\end{eqnarray}

\noindent the equation system(\ref{sys1}-\ref{sys4}) could be
reexpressed as the following autonomous system:

\begin{equation}\label{auto1}
\frac{dx}{dN}=\frac{3}{2}x[\gamma(1+x^2-y^2)-2x^2]-(3x+\sqrt{\frac{3}{2}}\lambda
y^2)
\end{equation}

\begin{equation}\label{auto2}
\frac{dy}{dN}=\frac{3}{2}y[\gamma(1+x^2-y^2)-2x^2]-\sqrt{\frac{3}{2}}\lambda
xy
\end{equation}

\noindent Also, we have a constraint equation

\begin{equation}\label{constraint}
\Omega_{\phi}+\frac{\kappa^2\rho_\gamma}{3H^2}=1
\end{equation}

\noindent where
\begin{equation}\label{omegar}
\Omega_{\phi}=\frac{\kappa^2\rho_{\phi}}{3H^2}=y^2-x^2
\end{equation}

The equation of state for the scalar fields could be expressed in
term of the new variables as

\begin{equation}\label{equaofstate}
 w_{\phi}=\frac{P_{\phi}}{\rho_{\phi}}=\frac{x^2+y^2}{x^2-y^2}
\end{equation}

\noindent It is not difficult to find the physically meaningful
critical points $(x_c, y_c)$ are $(0, 0)$ and
$(-\frac{\lambda}{\sqrt{6}}, \sqrt{1+\frac{\lambda^2}{6}})$. To
gain some insight into the property of the critical points, we
write the variables near the critical points $(x_c, y_c)$ in the
form

\begin{eqnarray}\label{stable}
x=x_c+u\\\nonumber y=y_c+v
\end{eqnarray}

\noindent where $u,v$ are perturbations of the variables near the
critical points. Substitute the expression (\ref{stable}) into the
autonomous system (\ref{sys1}-\ref{sys4}), one can obtain the
equations for the linear perturbations up to the first order as
following:

\begin{eqnarray}\label{perturbation}
\frac{du}{d N}=M_{11}u+M_{12}v\\\nonumber \frac{du}{d
N}=M_{21}u+M_{22}v
\end{eqnarray}

\noindent The coefficients of the perturbation equations form a
$2\times 2$ matrix $M$ whose eigenvalues determine the type and
stability of the critical points. One can easily find that the
eigenvalues of $M$ for $(0, 0)$ is $(\frac{3}{2}(-2+\gamma),
\frac{3\gamma}{2})$ and those for $(-\frac{\lambda}{\sqrt{6}},
\sqrt{1+\frac{\lambda^2}{6}})$ is $(-3-\frac{\lambda^2}{2},
-3\gamma-\lambda^2)$. Therefore, point (0,0) is a saddle point
while $(-\frac{\lambda}{\sqrt{6}}, \sqrt{1+\frac{\lambda^2}{6}})$
is a stable node, which corresponds to an attractor solution. At
this attractor solution, from Eq.(\ref{omegar}, \ref{equaofstate}
), we know that $\Omega_{\phi}=1$ and
$w_{\phi}=-1-\frac{\lambda^2}{3}$, which indicate the phantom
domination and the possibility of $w<-1$.

\begin{figure}
\psfig{file=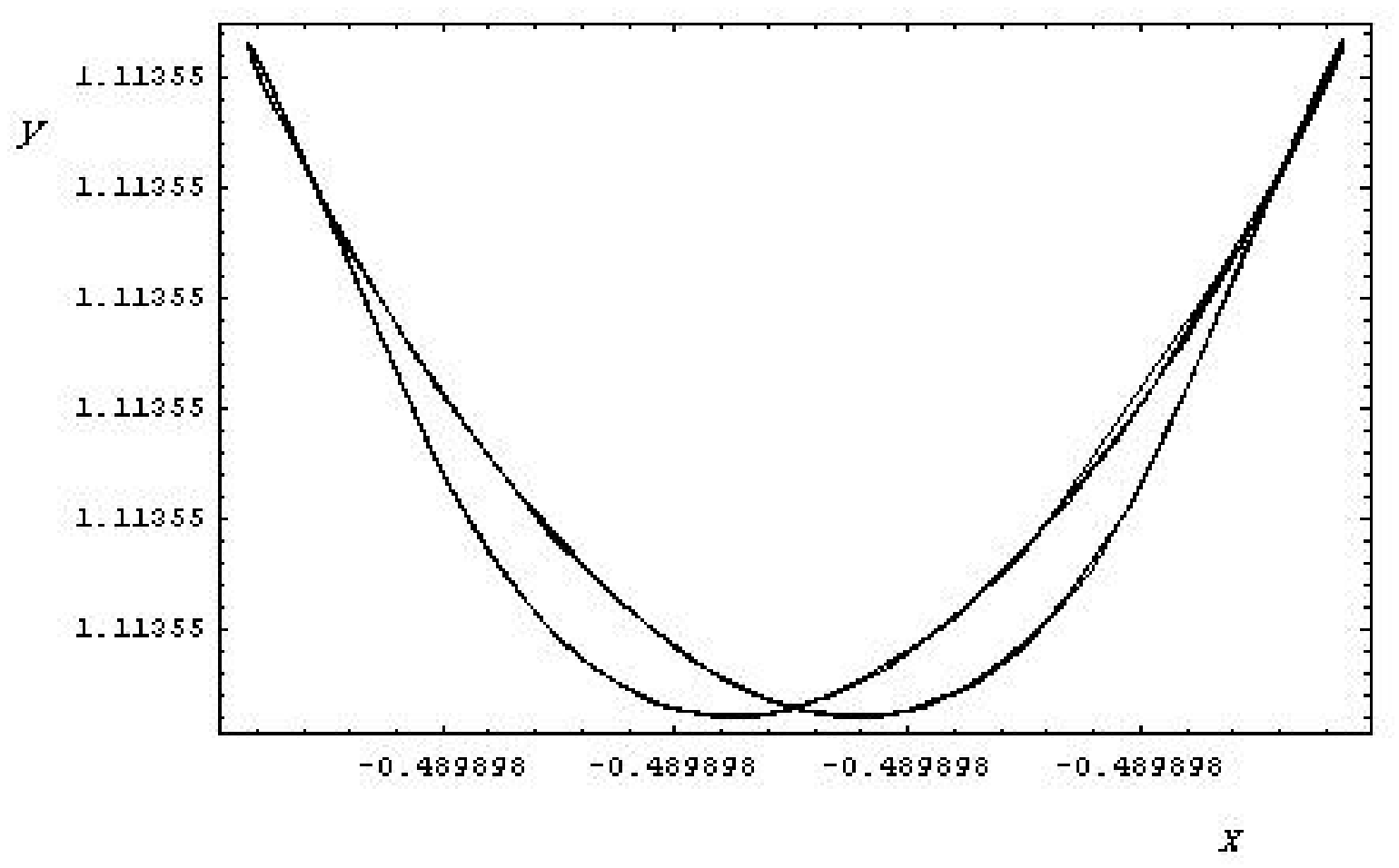,height=4in,width=5in} \caption{ the phase
graph of $y$ to $x$} \psfig{file=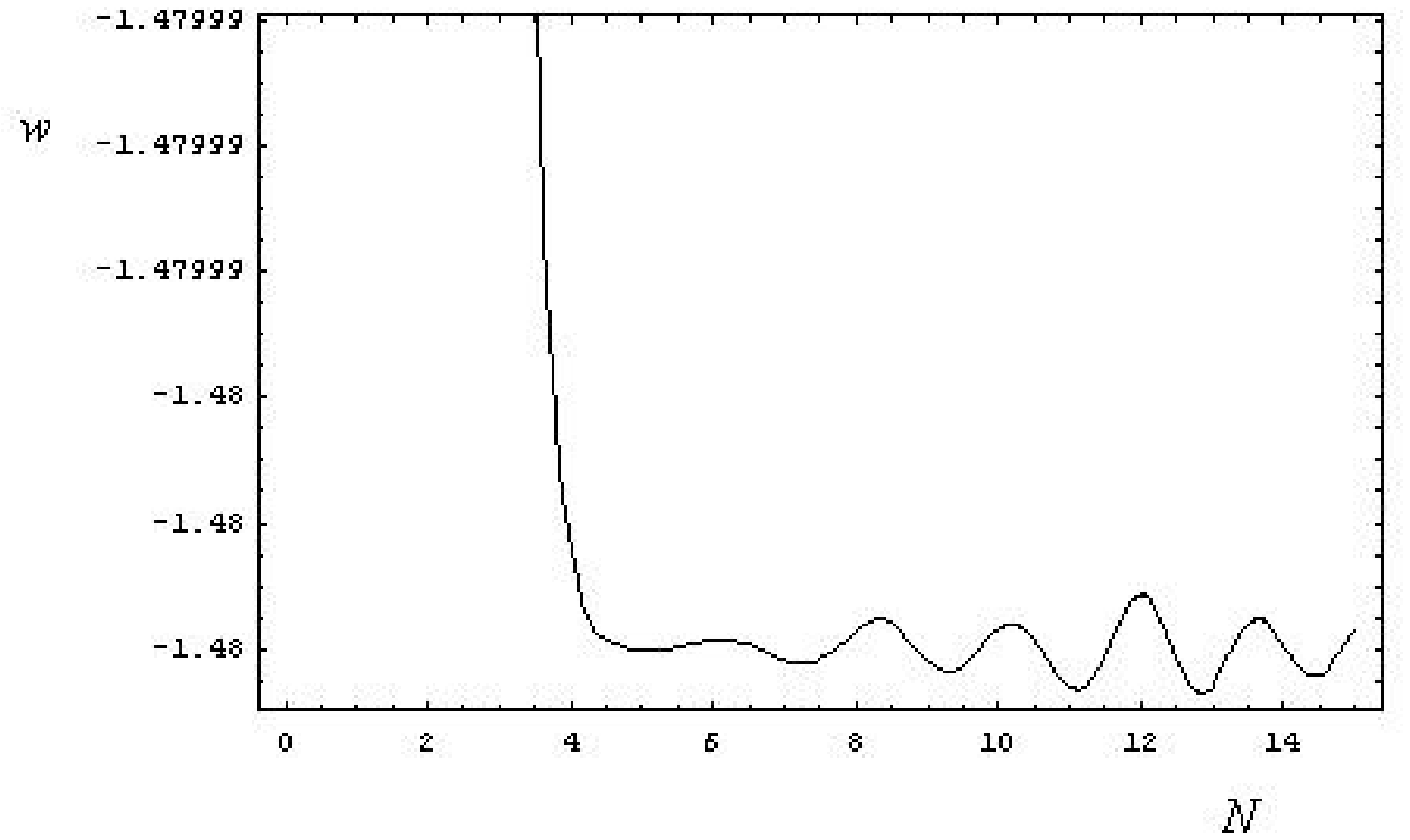,height=4in,width=5in}
\caption{ The evolution of Equation of state of Phantom $w$ with
respect to $N$}
\end{figure}
\begin{figure} \psfig{file=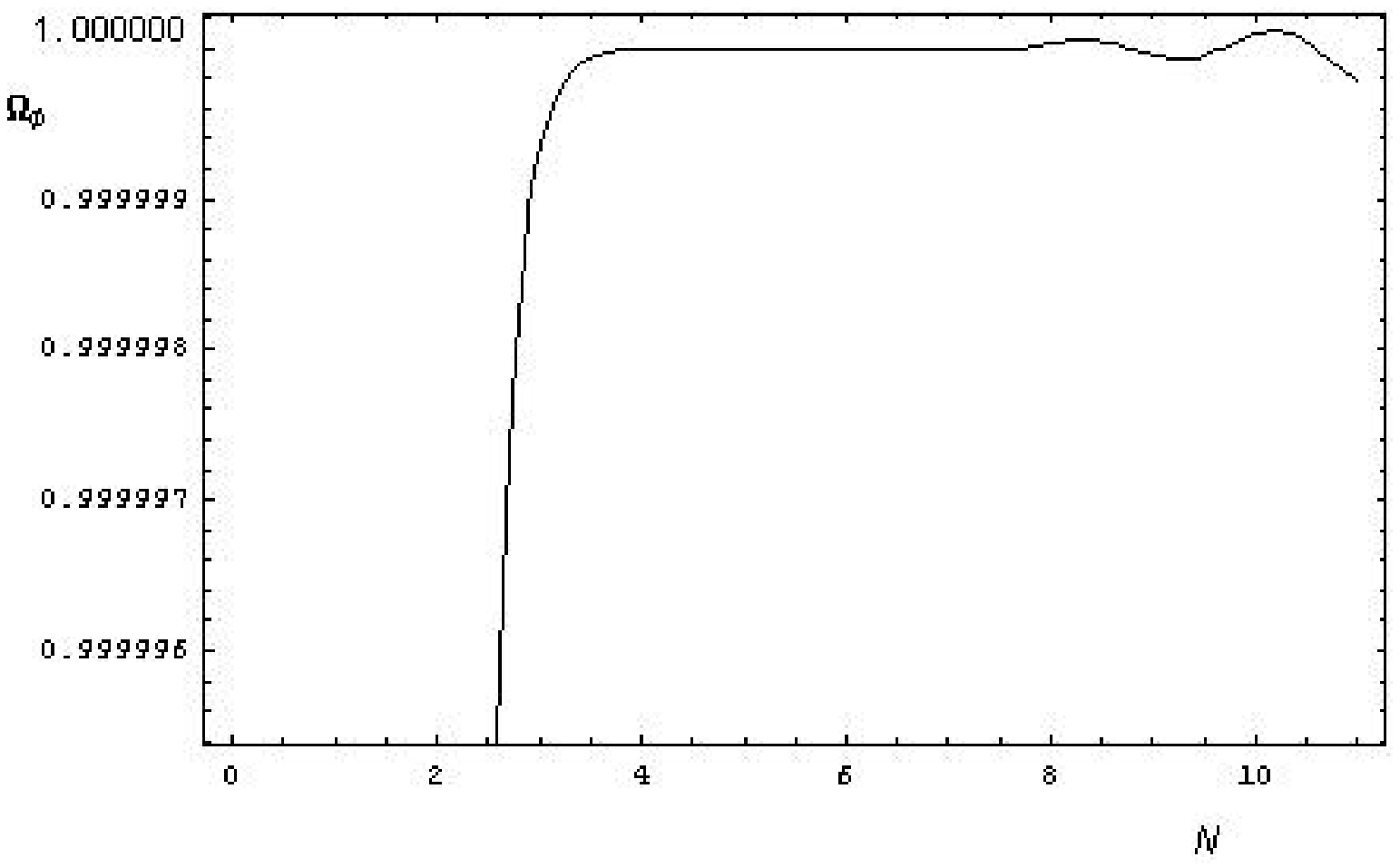,height=4in,width=5in}
\caption{The evolution of $\Omega_\phi$ with respect to $N $}
\end{figure}

Next, we study the system by numerical analysis. The results are
in Fig.1, Fig.2 and Fig.3. The computation was done at
$\lambda=1.2$ and $\gamma=1.0$. From the figures, we can find that
the attractor property of the solution and the phantom field will
slightly oscillate around the fixed point at late time and the
equation of state is smaller than $-1$.

\vspace{0.4cm} \noindent\textbf{4. Discussion} \vspace{0.4cm}

We studied a specific phantom model and show that there exists a
late time attractor solution in the evolution of the field. The
late time attractor solution corresponds to the phantom dominate
phase and the equation of state could be smaller than $-1$. This
is a very interesting feature of the model we considered here.
Now, we consider the stability of the phantom field under
perturbation in such a potential. The perturbed metric in
synchronous gauge could be expressed as\cite{caldwell1,carroll}

\begin{equation}\label{met}
ds^2=dt^2-a^2(t)(\delta_{ij}-h_{ij})dx^{i}dx^{j}
\end{equation}

\noindent Then, a Fourier mode of the phantom field
\begin{equation}\label{fourier}
\delta\phi(t,\textbf{k})=\frac{1}{\sqrt{2\pi}}\int
\delta\phi(t,\textbf{x})e^{-i\textbf{k}\cdot\textbf{x}}d^3x
\end{equation}

\noindent satisfies the equation of motion

\begin{equation}\label{perturbationeq}
\delta \ddot{\phi_k}+3H\delta
\dot{\phi_k}+(k^2-V"(\phi))=-\frac{1}{2}\dot{h}\dot{\phi}
\end{equation}

\noindent where the $h$ is the trace of $h_{ij}$ and the prime
denote the derivative with respect to $\phi$. The effective mass
for the perturbation is $[k^2-V"(\phi)]^{1/2}$. When the potential
is chosen as the exponential $V(\phi)=V_0\exp(-\lambda\kappa
\phi)$, we have $[k^2-\lambda^2\kappa^2V_0\exp(-\lambda\kappa
\phi)]^{1/2}$. When the field evolves to its stable attractor
solution, the cutoff wave number of the perturbation should be
\begin{equation}\label{conditionus}
k<\lambda H\sqrt{3+\frac{\lambda^2}{2}}
\end{equation}
\noindent so that the instability does not appear. Surely, it is
not to say that the instability will not appear during the
evolution before the field reaches the attractor. Since we do not
know what will be the specific evolution track the field took in
the past, we can only say that at present situation, which is the
attractor solution epoch, the perturbation should not violate
Eq.(\ref{conditionus}) to safeguard the stability of the phantom
field. Another remarkable feature of such an model is that the
evolution of field is toward to its attractor no matter what is
its initial value. Since $w_{\phi}=-1-\frac{\lambda^2}{3}$, it is
determined by the parameter $\lambda$ of the potential and is
independent of the choice of the initial value of the field, which
make the fine tuning of the field unnecessary.

\vspace{0.8cm} \noindent ACKNOWLEDGEMENT: This work was partially
supported by National Nature Science Foundation of China under
Grant No. 19875016, National Doctor Foundation of China under
Grant No. 1999025110, and Foundation of Shanghai Development for
Science and Technology under Grant No.01JC14035.

\end{document}